# A Kolmogorov-Zurbenko Fourier Transform Band-pass Filter Extension for Time Series Analysis

Edward Valachovic (1)*

(1) Department of Epidemiology and Biostatistics, College of Integrated Health Sciences, University at Albany, State University of New York, One University Place, Rensselaer, New York, United States of America

* Corresponding author
E-mail: evalachovic@albany.edu (EV)

## Abstract

This research introduces a novel extension, called the Extended Kolmogorov-Zurbenko Fourier Transform (EKZFT), to an existing class of band-pass filters first introduced by Kolmogorov and Zurbenko. Their original Kolmogorov-Zurbenko Fourier Transform (KZFT) is a useful tool in time series and spatio-temporal analysis, Fourier analysis, and related statistical analysis fields. Example uses of these filters include separating frequencies, filtering portions of the spectra, reconstructing seasonality, investigating periodic signals, and reducing noise. KZFT filters have many practical applications across a wide range of disciplines including health, social, natural, and physical sciences. KZFT filters are band-pass filters defined by three arguments: the length of the filter window; the number of iterations; and the central frequency of the band-pass filter. However, the KZFT filter is limited in design to only positive odd integer widow lengths inherited from the time series. Therefore, for any combination of the other KZFT filter arguments, there is only a relatively small, discrete, selection of possible filter window lengths in a range determined by the size of the dataset. This limits the utility of KZFT filters for many of the stated uses. The proposed EKZFT filter allows a continuous selection of filter window length arguments over the same range, offering improved control, increased functionality, and wider practical use of this band-pass filter. An example application of the EKZFT in a data simulation is provided.

# Introduction

In time series and spatio-temporal analysis, data observations are typically determined by the combined effects of several components such as trends, periodic components, random noise or variation, and other contributing factors. Whether for the purposes of investigation as the principal subject, or for exclusion from research into other component factors, periodic variation that occurs at a certain frequency, or the reciprocal of period, in time series are frequently a key focal point of research. Band-pass filters are a useful tool to help identify, separate, extract, and reconstruct variation in a time series that operates with a periodic pattern at a certain frequency. More on time series analysis and filters including definitions, notation, and examples can be found in Wei in addition to Shumway and Stoffer [1,2]. A band-pass filter, as the name implies, is designed so that the filtration process passes the energy or signal at frequencies present within a time series dataset that occur within a certain range or band of the frequency domain. The filter suppresses or attenuates frequencies that lie outside of that band.

The Kolmogorov-Zurbenko (KZ) filter, which is defined as an iteration of a central simple moving average filter, belongs to a class of low-pass filters that passes low frequencies and supresses or stops higher frequencies from passing through the filter and was first introduced by Zurbenko [3]. With notation $KZ_{m,k}$ the KZ filter has the two arguments $m$, the positive odd integer filter window length or size, and $k$, the positive integer number of iterations. An extension of the KZ filter, called the Kolmogorov-Zurbenko Fourier Transform (KZFT) is defined as an iteration of the Fourier Transform in Yang and Zurbenko [4]. The KZFT is effective as a band-pass filter, with a simple implementation of three functional arguments each with a clear interpretation. With notation $KZFT_{m,k,v}$ the arguments of the filter are $m$, the positive odd integer size or length of the filter window used in the filter calculations, $k$, the positive integer number of iterations of the Fourier transform, and $v$, the frequency at which the band-pass filter is centered. As such, the KZ filters are a special subset of KZFT filters when $v = 0$, in which case the band-pass is centered at a frequency of zero and it acts as a low-pass filter.

KZFT filters, as well as their low-pass KZ counterparts, have been used directly in a wide range of practical research applications, as well as indirectly as tools in other statistical methodologies. These filters have a history of use for time series analysis in a variety of fields across the physical, environmental, social, and health sciences. They were used extensively in meteorology and climatology in Wise and Comrie, and Gao et al. [5,6]. KZ filters were used for signal separation in studies of air quality in Kang et al. and Sezen et al. [7,8]. Zhang, Ma, and Kim used the filters to research environmental pollution and De Jongh, Verhoest, and De Troch used them to identify precipitation patterns [9,10]. For examples of their use as a tool in other methodologies, Valachovic and Zurbenko used these filters for filtration, separation and reconstruction of frequencies to identify patterns in skin cancer data and perform multivariate analysis on the component factors [11]. Recently, KZFT filters were used by Valachovic to design a new block bootstrapping method to preserve periodically correlated (PC) components in time series called the Variable Band-pass Periodic Block Bootstrap (VBPBB) [12]. In an extension of the VBPBB, filtration with multiple KZFT filters were instrumental in the design of a first of its kind method for resampling multiple periodically correlated (MPC) components in time series called the Variable Multiple Band-pass Periodic Block Bootstrap (VMBPBB) [13].

While many of these examples emphasise the use of KZFT filters in analysis and as a tool of other methodologies to reduce the influence of interfering frequencies, supress random noise, and separate and filter portions of the frequency domain prior to analysis, they also exposed shortcomings with the KZFT. Like the KZ filter, the KZFT filter has the argument for the filter window length restricted to positive odd integer values. This may be sufficient for some applications and purposes, such as creating a filter to separate time series components by frequency when the target frequencies are widely separated in the frequency domain. However, certain applications such as the VBPBB and VMBPBB were shown to benefit from fine adjustment of the band-pass filters used in their design. Particularly when the filter argument for the window length is relatively small, the set of available KZFT filters and their correspond band-pass frequency regions differ greatly. This can be problematic for these methods when trying to separate frequencies that are very close together. There is a greater need for fine adjustment of the band of frequencies that are passed particularly regarding frequencies that are excluded and supressed or attenuated, and this adjustment may need to be performed using only the argument for the filter window length due to data limitations.

Recently, a novel extension of the KZ filter was proposed by Valachovic permitting iterated central simple moving averages to be calculated on any real valued window size or length, unlike the design of the KZ filter which is restricted to positive odd inter valued filter windows [14]. This Extended Kolmogorov-Zurbenko (EKZ) filter was shown share the properties of the KZ filters when the window size is a positive odd integer since the KZ filters comprised a set of special cases of the EKZ filter and are mathematically equivalent to EKZ filters in these circumstances. Where the EKZ filter differed was in its design which permitted a continuous range of window size arguments, over intervals where the KZ filter only permitted a limited finite set of arguments. The KZFT filter shares the same shortcoming as KZ filters with respect to restrictions on the filter window length, therefore the research aim of this work is to adapt the new methods designed within the EKZ filter to create an extension of the KZFT.

An Extended Kolmogorov-Zurbenko Fourier Transform (EKZFT) contains the KZFT filters as a set of special cases of the EKZFT. When the window length is a positive odd integer, they are mathematically equivalent and therefore share the properties of KZFT filters in those circumstances. Furthermore, like the EKZ filter, the EKZFT filter properties are inherited from the KZFT filter even when window length sizes are values that are not defined for the KZFT filter. This is particularly helpful in optimization of the VBPBB and VMBPBB methods which require fine adjustment of the band-pass filter arguments.

The shortcomings of the KZFT band-pass filter and the need for the proposed EKZFT filter can best be understood when visualizing certain properties of the KZFT. Yang and Zurbenko provide the energy transfer function for the KZFT filter, which is a mapping that describes how input frequencies are transferred to outputs [4]. Additional information on energy transfer functions and the energy transfer functions of KZFT filters can be found in Shumway and Stoffer in addition to Zurbenko [2,3]. The energy transfer function is useful to show the effect of a KZFT filter on a time series and how with only a few iterations this class of filters strongly attenuates signals with frequency shift or separation of $1/m$ and farther away from the central frequency $v$, creating the band of frequencies that are passed by the filter. Fig 1 illustrates examples of KZFT filters, all with a fixed argument of $m = 7$ and centered at a frequency $v$, but paired with different choices for $k$, the argument of the number of iterations. The energy transfer functions show the energy transfer at different frequency shifts away from $v$, the argument for the central frequency for the band, for $k = 1$ is in magenta, for $k = 2$ in red, $k = 3$ in yellow, $k = 4$ in green, $k = 5$ in cyan, and $k = 6$ in blue. Fig 1 to the left shows that the attenuation is complete at a frequency shift $1/m$ = 1/7 and its harmonics, or integer multiples of this frequency, so 2/7, 3/7, and so on. However, the $KZFT_{m=7,k=1,v}$ in magenta only modestly suppresses other frequencies shifted farther than $1/m$ away from $v$. The advantage of higher iterations of the KZFT filters is that higher iterations improve this attenuation or suppression outside of the band of frequencies passed. At two iterations in red, the $KZFT_{m=7,k=2,v}$ more completely suppresses these other frequencies, but looking closely there is still visible evidence of leakage through the filter at farther frequency separation. At $k = 3$ and above, in Fig 1 to the left, the energy transfer function does not appear to be visibly different from zero at frequencies shifts greater than $1/m$. However, in Fig 1 to the right the natural logarithm of the energy transfer functions reveals that this is not quite the case. As the number of iterations increases the KZFT filters completely attenuate at a frequency shift of $1/m$ and its harmonics, and the degree of attenuation at frequency shifts greater than $1/m$ becomes more complete.

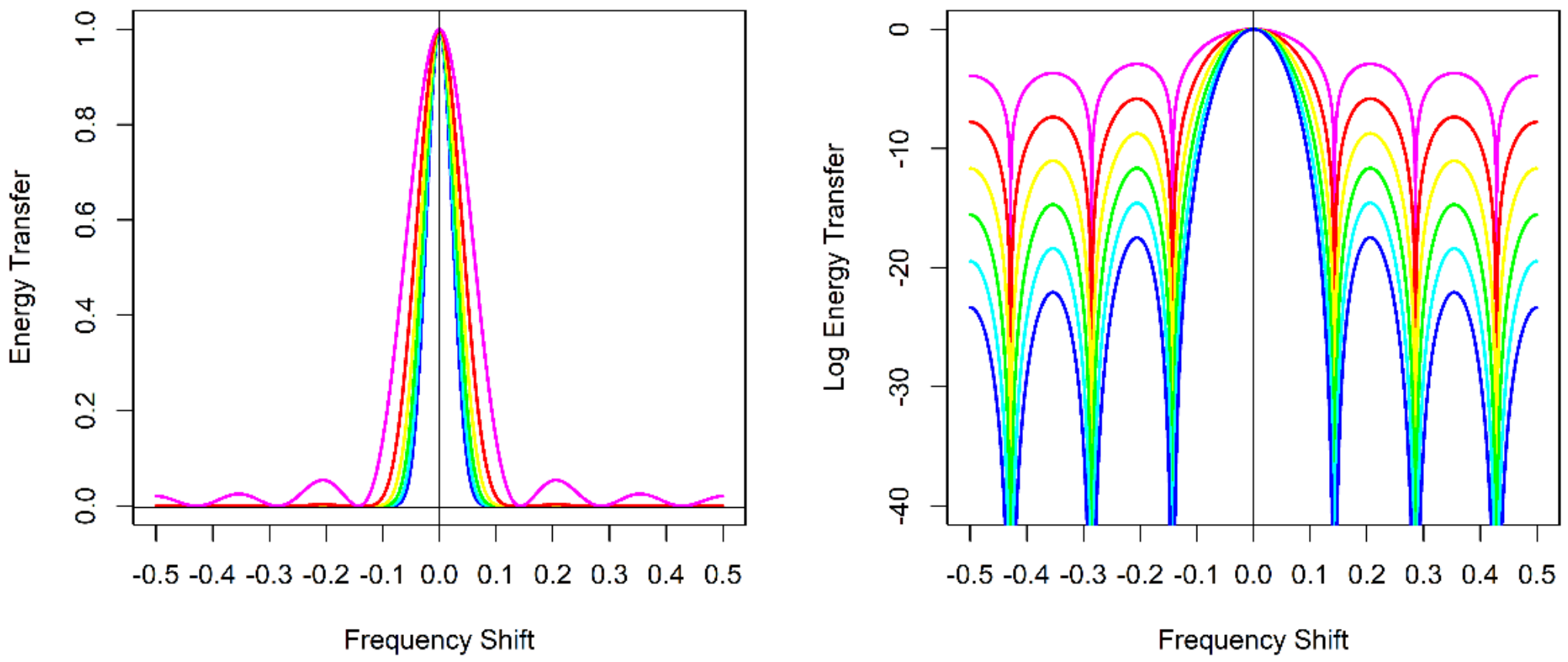

**Fig 1. Kolmogorov-Zurbenko Fourier Transform filter energy and log energy transfer functions.** The energy transfer functions at different frequency shifts to the left and log energy transfer function to the right for Kolmogorov-Zurbenko filters with a fixed argument of $m = 7$ paired with $k = 1$ in magenta, $k = 2$ in red, $k = 3$ in yellow, $k = 4$ in green, $k = 5$ in cyan, and $k = 6$ in blue.

However, the design shortcoming of the KZFT filter is its constraint that the selection of filter window lengths must be positive odd integers for the argument $m$. The limitation of this constraint becomes apparent with the energy transfer function. The constraint on the filter window length is severely limiting in the choice of which frequency shift that can be completely suppressed. With this limitation, there are continuous interval gaps regarding what frequency shift away from the central band-pass frequency that cannot be completely suppressed. Fig 2 shows this shortcoming in the energy transfer functions for KZFT filters with different frequency shifts away from $v$, with a fixed argument of $k = 1$ and paired with $m = 3$ in red, $m = 5$ in green, $m = 7$ in blue, $m = 9$ in dashed red, $m = 11$ in dashed green, $m = 13$ in dashed blue. Clearly, as $m$ increases, the band of frequencies that the energy transfer function passes shrinks and crops near the central frequency, $v$, corresponding to zero frequency shift. Also, the frequency shift $1/m$ which is completely attenuated also moves toward zero. As a result, the design options of KZFT filters are sparse at relatively small values for the argument $m$ as compared to higher values of $m$. Seen in Fig 2, there are only four choices for the filter window lengths in KZFT filters between the frequency separations of 0.1 and 0.5.

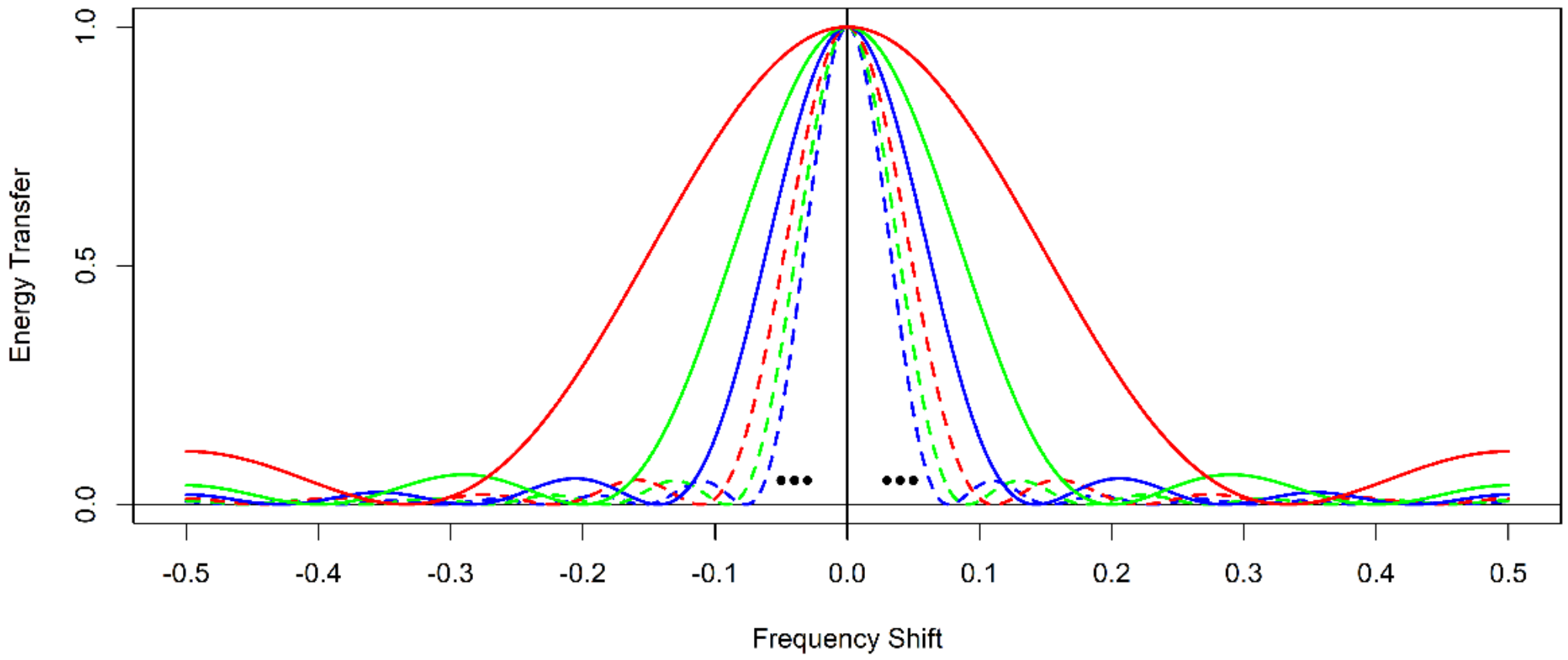

**Fig 2. Kolmogorov-Zurbenko Fourier Transform energy transfer functions for different values of argument *m* with fixed argument *k*.** The energy transfer function at different frequency shifts for Kolmogorov-Zurbenko filters with a fixed argument of $k = 1$ paired with $m = 3$ in red, $m = 5$ in green, $m = 7$ in blue, $m = 9$ in dashed red, $m = 11$ in dashed green, $m = 13$ in dashed blue. The ellipses shows that there are additional KZFT energy transfer functions in this region as $m$ increases.

This is very limiting when designing KZFT filters for signal separation and reconstruction. For instance, it may be the case that two component signals have periods that differ from each other by a positive even integer, and the goal is to pass one component and suppress the other. While it is possible to approximate a filter that completely passes one while fully suppressing the other using a KZFT filter designed with an odd positive filter window length close to that positive even integer, the filtration will be incomplete. Furthermore, as Fig 2 illustrated, the approximation is increasingly poor for smaller values of $m$. For another example, if the goal is separating harmonics of a given fundamental frequency, those harmonics are likely to have non-integer period separations with the fundamental frequency which can not be designed into the KZFT filter window length.

The EKZFT filter, introduced here, is designed to solve this limitation in the KZFT filter. Borrowing from the EKZ, the EKZFT is an iteration of a Fourier transform so that it permits any positive real valued window length and is formally defined in the next section. To accomplish this, it treats a time series, with its original unit time measure, as if it was recorded over intervals of smaller time units, or even continuously. Since any positive real valued number greater than one, $m_r$, can be expressed as the sum of the closest positive odd integer less than that real number, $m_o$, and the difference, $m_d = m_r - m_o$, that difference must be less than two. The EKZFT design with a filter window size that is not a positive odd integer will add two extra terms to the calculated average performed as part of the KZFT. Compared to the KZFT filter with the next lower positive odd integer, $m_o$, the EKZFT adds the next term preceding and the next term following the $m_o$ ordered summed terms, and those two additional terms are each weighted by one half the difference, $m_d/2 = m_r - m_o/2$. Therefore, any positive real valued window length, $m_r$, can be used to create an EKZFT formed from the ordered values of the original time series, with weights 1 for the nearest $m_o$ observations surrounding the central time point, and weights equal to half of $m_d$ for the next farther observations before and after. With this design, the EKZFT is identical to a KZFT filter if it had it been applied to the time series were measured in finer time units than it is as given. Therefore, the EKZFT is an extension of the KZFT filter, and it retains many properties of the KZ filter, while permitting a continuous choice of filter window length, $m_r$.

# Methods

## The Kolmogorov-Zurbenko Fourier Transform

The original Kolmogorov-Zurbenko Fourier Transform (KZFT) is the iteration of the conventional Fourier transform defined in Yang and Zurbenko [4]. It can be treated as a linear band-pass filter with three arguments $m$, $k$, and $v$ and notation $KZFT_{m,k,v}$. The argument $m$, a positive odd integer, is the length of the KZFT filter window, the argument $k$, a positive integer, is the number of iterations of the Fourier transform, and the frequency $v$ is the center of the band-pass filter. Note, the $KZ_{m,k}$ filter which is a central moving average filter of window length $m$ iterated $k$ times, is a special case of the $KZFT_{m,k,v}$ filter when $v = 0$ [4]. As a band-pass filter, the KZFT completely attenuates signals at a frequency shift of $1/m$ and its integer multiples or harmonics from the central frequency $v$. Furthermore, it strongly attenuates, though not completely attenuates, all other signals at a frequency shift of $1/m$ and higher while passing frequencies closer to the central frequency $v$. The effect of a $KZFT_{m,k,v}$ filter is preserving periodic variation in a band of frequencies near *v* while smoothing variation outside that band of frequencies. The equations below follows the notation found in Zurbenko, and Yang and Zurbenko [3,4]. Applied to a time series process $\{X(t): t \in \mathbb{Z}\}$, a KZFT filter of *m* data-points and *k* iterations, centered at frequency *v*, where *t* is time ordering the data-points, *s* are the time steps defining the symmetric centered filter window, and $a_s^{m,k}$ are coefficient weights, is defined as follows:

$$KZFT_{m,k,v}\big(X(t)\big) = \sum_{s=-\frac{k(m-1)}{2}}^{\frac{k(m-1)}{2}} \frac{a_s^{m,k}}{m^k} e^{-i2mvs} X(t+s) \tag{1}$$

The coefficients $a_s^{m,k}$ can be conveniently found as the polynomial coefficients obtained from an expansion of the following polynomial (arbitrarily in *z*):

$$\sum_{s=0}^{k(m-1)} z^s a_{s-k(m-1)/2}^{m,k} = (1 + z + \cdots + z^{m-1})^k \tag{2}$$

The coefficients are obtained by the convolution of k uniform discrete distributions on the interval $\left[-\frac{m-1}{2}, \frac{m-1}{2}\right]$ where *m* is an odd integer. Therefore, each coefficient forms a tapering window which has finite support. For some examples calculating the coefficients, a $KZFT_{m=5,k=1,v=0.25}$ filter would have an ordered sequence of $a_s^{5,1}$ coefficients {1,1,1,1,1}, and a $KZFT_{m=5,k=2,v=0.25}$ filter would have an ordered sequence of $a_s^{5,2}$ coefficients {1,2,3,4,5,4,3,2,1}.

The KZFT filter can be produced with statistical software like R Studio with packages such as KZA detailed in Close and Zurbenko [15,16]. The KZFT filter can also be obtained in an iterated form according to the following algorithm found in Yang and Zurbenko [4]. In the iterated form of the $KZFT_{m,k,v}$ filter each step is an application of a $KZFT_{m,k=1,v}$ filter to the prior result thus making all $a_s^{m,1}$ coefficients equal to one in each step. The iterated form of the $KZFT_{m,k,v}$ is provided in the following equations:

$$KZFT_{m,1,v}\big(X(t)\big) = \sum_{s=-\frac{m-1}{2}}^{\frac{m-1}{2}} \frac{a_s^{m,1}}{m^1} e^{-i2mvs} X(t+s)$$

$$= \frac{1}{m} \sum_{s=-\frac{m-1}{2}}^{\frac{m-1}{2}} e^{-i2mvs} X(t+s) \tag{3}$$

$$KZFT_{m,2,v}\big(X(t)\big) = \frac{1}{m} \sum_{s=-\frac{(m-1)}{2}}^{\frac{(m-1)}{2}} KZFT_{m,1,v}\big(X(t+s)\big) \tag{4}$$

$$\vdots$$

$$KZFT_{m,k,v}\big(X(t)\big) = \frac{1}{m}\sum_{s=-\frac{(m-1)}{2}}^{\frac{(m-1)}{2}} KZ_{m,k-1,v}\big(X(t+s)\big) \tag{5}$$

One tool that characterizes the effect of a KZFT filter is the energy transfer function, or square of the linear mapping, *B*, that describes how input frequencies are transferred to outputs. Zurbenko and separately Yang and Zurbenko introduce the energy transfer function for the KZFT filter [3,4]. The energy transfer function is useful to show the effect of a KZFT filter on a time series, and how with only a few iterations, this filter strongly attenuates signals with a frequency shift $1/m$ and farther away from $v$ while passing frequencies within the band around $v$. The strength of that suppression, or attenuation, is complete with a frequency shift $1/m$ and its multiples away from $v$. The energy transfer function of the KZFT filter at each frequency $\lambda$ is seen in the following equation:

$$|B(\lambda - v)|^2 = \left(\frac{\sin(\pi m(\lambda - v))}{m\sin(\pi(\lambda - v))}\right)^{2k} \tag{6}$$

Since the KZFT energy transfer function characterizes how inputs are passed verses supressed, it is also useful to define the cut-off frequency which is a defined limit or boundary at which the energy transferred through a filter is generally suppressed or diminished rather than allowed to pass through. A cut-off frequency where output power is half that of the input, called the half power point, for the KZFT filter is provided by Zurbenko, and Yang and Zurbenko, below [3,4].

$$|\lambda_0 - \mathrm{v}| \approx \frac{\sqrt{6}}{\pi}\sqrt{\frac{1-\left(\frac{1}{2}\right)^{\frac{1}{2k}}}{m^2-\left(\frac{1}{2}\right)^{\frac{1}{2k}}}} \tag{7}$$

# The Extended Kolmogorov-Zurbenko Fourier Transform

The proposed extension to the KZFT, called the Extended Kolmogorov-Zurbenko Fourier Transform (EKZFT), is equivalent to an extension of the iterated Fourier Transform. The design of the EKZFT originates from the Extended Kolmogorov-Zurbenko (EKZ) filter, which extended the original Kolmogorov-Zurbenko (KZ) filter defined as the iteration of a central simple moving average (SMA). Just as the KZ filter is a low-pass filter and the KZFT is a band-pass filter equivalent to a KZ filter centered at a frequency *v*, the EKZFT is an EKZ filter centered at a frequency *v*. Similarly, whereas the KZ filters are a subset of special cases of the EKZ, so too are the KZFT filters which are a subset of special cases of the EKZFT filters. The EKZ low-pass filter developed from the aim to define a central SMA or KZ filter over finer time units than those measured by a given time series for signals that are not integer multiples of those time units. The design of a central SMA filter averages all data observations with equal weight in the filter window length. Similarly, the KZ filter balances the number of observations before, and after which centralizes the filter and results in the limitation to only positive odd integer window lengths. Rather than equally weighting observations, in order to define a central equally weighted moving average for numbers other than positive odd integer window lengths, the EKZ creates an extension of the KZ filter by iterating a specially, and unequally, weighted central moving average filter on a positive odd integer window length. In effect, the EKZ filter creates moving averages that are equal to what they would be with KZ filters, were the data recorded finely enough so that the desired window length was a positive odd integer. This permits the EKZ to be applied as if it were an SMA applied to any window length, including those that are not positive odd integer lengths. The design of the EKZFT adapts this idea to the iterated Fourier transform or KZFT by creating a Fourier transform equal to what they would be with KZFT filters, were the data recorded finely enough so that the desired window length was a positive odd integer. Therefore, it is a filter with notation $EKZFT_{m_r,k,v}$ and with three arguments $m_r$, $k$, and $v$. The argument $m_r$ is any positive real number greater than one for the filter window length. The argument $k$ remains a positive integer for the number of iterations, and the argument $v$ remains the frequency of the center of the band-pass filter. As noted previously, the KZFT filters are a subset of special cases of EKZFT filters, so an $EKZFT_{m_r,k,v}$ filter is equivalent to a $KZFT_{m,k,v}$ filter when $m_r = m$, a positive integer greater than one.

To define the EKZFT filter, let $\{X(t): t \in \mathbb{Z}\}$ be a time series where $t$ is time. For an $EKZFT_{m_r,k,v}$ filter of a positive real valued filter window length of $m_r > 1$ time points and positive integer $k$ iterations, define $m_r = m_o + m_d$ time points where $m_o$ is the nearest positive odd integer less than $m_r$, and $m_d$ is the difference equal to $m_r - m_o$. Clearly, it must be that $0 \le m_d < 2$. Then an $EKZFT_{m_r,k,v}$ filter of $m_r$ time points and $k$ iterations, and centered at frequency $v$, where $s$ are the steps defining the symmetric filter window, and $a_s^{m,k}$ are coefficient weights, is defined as follows:

$$EKZFT_{m_r,k,v}\big(X(t)\big) = \frac{1}{(m_r)^k} \sum_{s=-\frac{k(m_o+1)}{2}}^{\frac{k(m_o+1)}{2}} a_s^{m_r,k} e^{-i2mvs} X(t+s) \qquad (8)$$

The coefficients $a_s^{m_r,k}$ can be conveniently found as the polynomial coefficients obtained from an expansion of the following polynomial (arbitrarily in $z$):

$$\sum_{s=0}^{k(m_o+1)} z^s a_{s-k(m_o+1)/2}^{m_r,k} = \left(\frac{m_d}{2} + z + \cdots + z^{m_o} + \frac{m_d}{2} z^{m_o+1}\right)^k \qquad (9)$$

The coefficients are obtained by the convolution of k uniform discrete distributions on the interval $\left[-\frac{m_o+1}{2}, \frac{m_0+1}{2}\right]$ where $m_o$ is an odd integer, and each coefficient forms a tapering window which has finite support. The difference in the EKZFT filter from the KZFT filter is that the first and last coefficients in the polynomial are $\frac{m_d}{2}$ rather than one. For some examples calculating the coefficients, an $EKZFT_{m_r=2.5,k=1,v=0.25}$ filter would have an ordered sequence of $a_s^{2.5,1}$ coefficients $\{0.75,1,0.75\}$. An $EKZFT_{m_r=2.5,k=2,v=0.25}$ filter would have an ordered sequence of $a_s^{1.5,2}$ coefficients $\{0.75,1.75,2.5,1.75,0.75\}$. Lastly, a $EKZFT_{m_r=\pi,k=1,v=0.25}$ would have ordered sequence of $a_s^{\pi,1}$ coefficients $\left\{\frac{\pi-3}{2}, 1,1,1, \frac{\pi-3}{2}\right\}$.

In the iterated form of the $EKZFT_{m_r,k,v}$ filter each step is an application of an $EKZFT_{m_r,k=1,v}$ filter to the prior result. In this form, since $k$=1 in each step, all $a_s^{m_r,1}$ coefficients are equal to one in each step except for the first and last coefficients provided $m_d \neq 0$. The iterated form of the $EKZFT_{m_{r_r},k,v}$ is provided in the following equations:

$$EKZFT_{m_r,1,v}\big(X(t)\big) = \frac{\frac{1}{2}m_d}{m_r} e^{-i2mv\left(-\frac{m_o+1}{2}\right)} X\left(t - \frac{m_o+1}{2}\right) + \left[\sum_{s=-\frac{m_o-1}{2}}^{\frac{m_o-1}{2}} \frac{1}{m_r} e^{-i2mvs} X(t+s)\right] + \frac{\frac{1}{2}m_d}{m_r} e^{-i2mv\left(\frac{m_o+1}{2}\right)} X\left(t + \frac{m_o+1}{2}\right) \quad (10)$$

$$EKZ_{m_r,2,v}\big(X(t)\big) = \frac{\frac{1}{2}m_d}{m_r} EKZFT_{m_r,1,v}\left(X\left(t - \frac{m_o+1}{2}\right)\right) + \left[\sum_{s=-\frac{m_o-1}{2}}^{\frac{m_o-1}{2}} \frac{1}{m_r} EKZFT_{m_r,1,v}\big(X(t+s)\big)\right] + \frac{\frac{1}{2}m_d}{m_r} EKZFT_{m_r,1,v}\left(X\left(t + \frac{m_o+1}{2}\right)\right) \quad (11)$$

$$\vdots$$

$$EKZFT_{m_r,k,v}\big(X(t)\big) = \frac{\frac{1}{2}m_d}{m_r} EKZFT_{m_r,k-1,v}\left(X\left(t - \frac{m_o+1}{2}\right)\right) + \left[\sum_{s=-\frac{m_o-1}{2}}^{\frac{m_o-1}{2}} \frac{a_s^{m,1}}{m_r} EKZFT_{m_r,k-1,v}\big(X(t+s)\big)\right] + \frac{\frac{1}{2}m_d}{m_r} EKZFT_{m_r,k-1,v}\left(X\left(t + \frac{m_o+1}{2}\right)\right) \quad (12)$$

The EKZFT filters inherit or are approximated by many of the properties of KZFT filters, because KZFT are readily seen as a special subset of EKZFT filters when argument $m_r = m$ as a positive odd integer. Using the same reasoning presented for the EKZ filter by Valachovic, the EKZFT filter properties are also characterized in three cases: when the window length is a positive odd integer, when it is a positive even integer, and when it is any other real number greater than one [14]. For EKZFT filters where $m_r = m$, a positive odd integer, their definition mathematically simplifies to that of KZFT filters. It is straightforward to see that all equations of the EKZFT simplify and are identical to that of the KZFT where $m_d = 0$, the consequence of $m_r = m$ as a positive odd integer. Consequently, just like the KZFT filter, the $EKZFT_{m_r=m,k,v}$ filter strongly attenuates signals with a frequency shift $1/m$ and farther away from $v$ while passing frequencies within the band around $v$. That suppression is complete with a frequency shift of precisely $1/m$ and its multiples or harmonics away from $v$. When $m_r$ is a positive even integer, a case that the KZFT filter can not accommodate, the EKZFT filter still strongly attenuates signals with a frequency shift $1/m_r$ and farther away from $v$ while passing frequencies within the band around $v$. That suppression is again complete with a frequency shift of precisely $1/m_r$ and its multiples away from $v$. However, when $m_r$ is any other real number greater than one, the properties of the EKZFT filter are only approximated by the properties of the KZFT filter. When $m_r$ is any other real number greater than one, the EKZFT still strongly attenuates signals with a frequency shift $1/m_r$ and farther away from $v$ while passing frequencies within the band around $v$, just as in the other scenarios. However, that suppression is no longer guaranteed to be complete with a frequency shift of precisely $1/m_r$ and its multiples away from $v$.

Therefore, the energy transfer function inherited from the KZFT filter is best considered as only an approximate for the EKZFT filter at each frequency $\lambda$. Specifically, the form of the KZFT energy transfer function is the same when $m_r$ is a positive odd value greater than one, and approximate for all other values greater than one, and is seen in the following equation:

$$|B(\lambda - v)|^2 = \left(\frac{\sin(\pi m_r(\lambda - v))}{m_r \sin(\pi(\lambda - v))}\right)^{2k} \tag{13}$$

Likewise, the cut-off frequency where output power is half that of the input, or the half power point, for the EKZFT filter energy transfer function is approximated by the equation provided below:

$$|\lambda_0 - \mathrm{v}| \approx \frac{\sqrt{6}}{\pi}\sqrt{\frac{1-\left(\frac{1}{2}\right)^{\frac{1}{2k}}}{m_r^2-\left(\frac{1}{2}\right)^{\frac{1}{2k}}}} \tag{14}$$

Fig 3 shows the energy transfer function for multiple EKZFT filters over a small range of filter window lengths, many of which are not otherwise possible with KZFT filters. To demonstrate flexibility, Fig 3 shows EKZFT filters with the number of iterations fixed at $k = 1$, frequency centered at $v$, and the moving average window length ranging from $m_r = 1$ to $m_r = 7$ by one-unit intervals. Between each interval, values of $m_r$ are used to illustrate the designs of filters possible. Of course, this is only illustrative of the range of EKZFT filters possible, of which there are an infinite number for any real number within this range. For comparison, there are only three KZFT filters in the same range where argument $m$ = 3, 5, or 7, and whose energy transfer functions are identified by the black dashed lines in Fig 3. Finally, more filters are possible as $m_r$ increases above 7, and the energy transfer function becomes more compact around the zero-frequency shift indicated by the ellipsis in Fig 3.

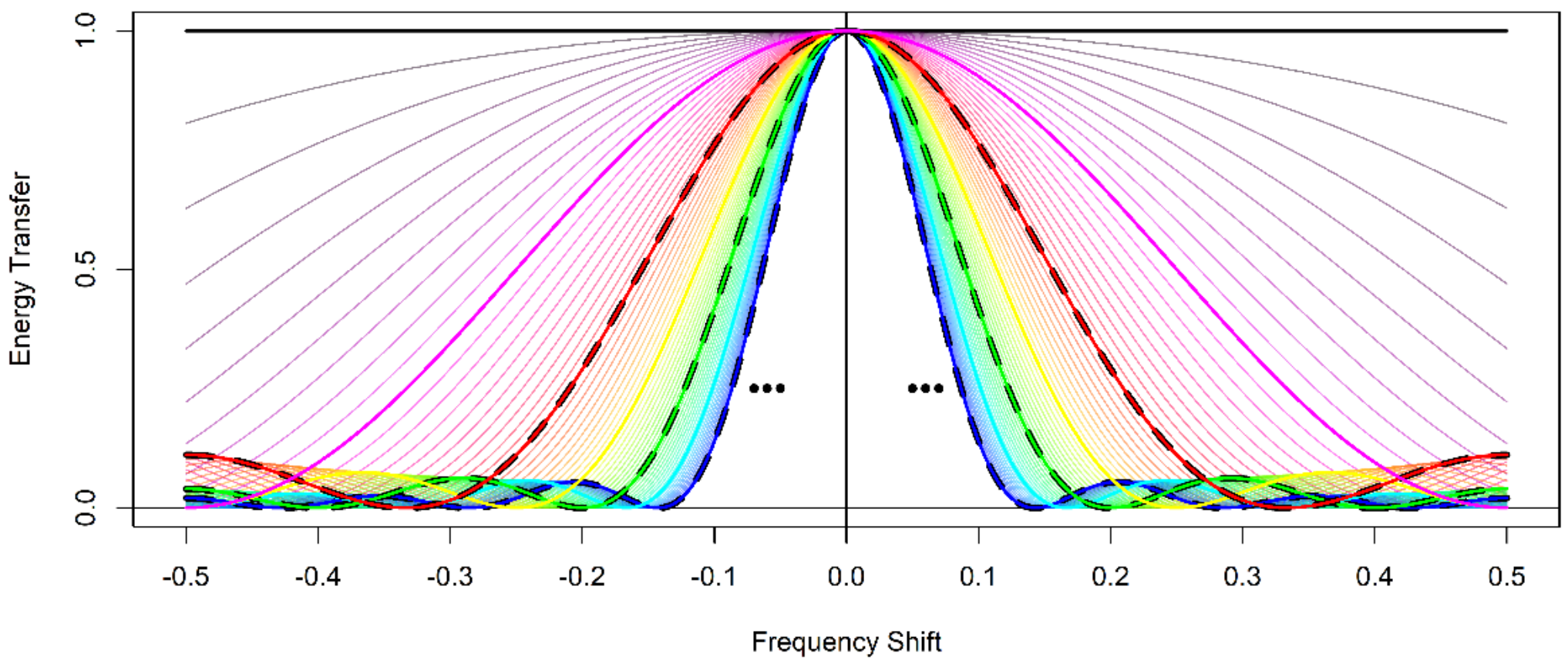

**Fig 3. Extended Kolmogorov-Zurbenko Fourier Transform filter energy transfer functions.** The energy transfer function at different frequency shifts for Extended Kolmogorov-Zurbenko Fourier Transform filters with arguments $k = 1$, centered at frequency $v$, and $m_r = 1$ in black, $m_r = 2$ in magenta, $m_r = 3$ in red, $m_r = 4$ in yellow, $m_r = 5$ in green, $m_r = 6$ in cyan, and $m_r = 7$ in blue, in addition to $m_r$ within each created interval. The lines with black dashes are the only possible energy transfer functions for KZ filters in the same range of argument $m$ at $m = 3$ in red, $m = 5$ in green, and $m = 7$ in blue.

# Simulation

To demonstrate the use of the EKZFT filter proposed here, as well as illustrate the results obtained, this work uses a simulated time series. In the simulation KZFT filters are applied to identify their limitation, and EKZFT filters are applied with arguments that are not possible with KZFT filters. Periodograms, which are described in Wei, and Shumway and Stoffer, and illustrate sample variance at each frequency within a time series, are used to observe the effects of different EKZFT and KZFT filters upon the original simulated time series [1,2].

Analysis is performed in R version 4.1.1 (2013) statistical software using the KZFT function in the KZA package, see Close and Zurbenko for more detail, with datasets as a time series measured on an ordered interval dimension, in this case time [15,16]. The simulated time series is constructed with 100,000-time units. First, a time series of random variation or white noise is generated with independent draws from a standard normal distribution with mean zero and standard deviation one. The periodogram of this white noise time series data is seen in Fig 4, plotted using R in black, and exhibits random low fluctuating levels of energy across all frequencies which is characteristic of white noise. The periodogram of an $EKZFT_{m_r=8,k=1,v=0.25}$ filter applied to this time series in magenta, exhibits the successful suppression of variation at a frequency shift of $1/m_r = 1/8$ and its harmonic $2/m_r = 2/8 = 1/4$ away from the central frequency $v$ = 0.25. This matches the energy transfer function of an $EKZFT_{m_r=8,k=1,v=0.25}$ filter plotted as the magenta curve with corresponding scale on the right. For comparison, the closest possible KZFT filters to the $EKZFT_{m_r=8,k=1,v=0.25}$ are the $KZFT_{m=7,k=1,v=0.25}$ with energy transfer function displayed in blue and $KZFT_{m=9,k=1,v=0.25}$ with energy transfer function displayed in red, both of which are clearly seen not suppress variation at a frequency shift of $1/m_r = 1/8$ away from the central frequency $v$ = 0.25, but rather at 1/7, and 1/9 away respectively.

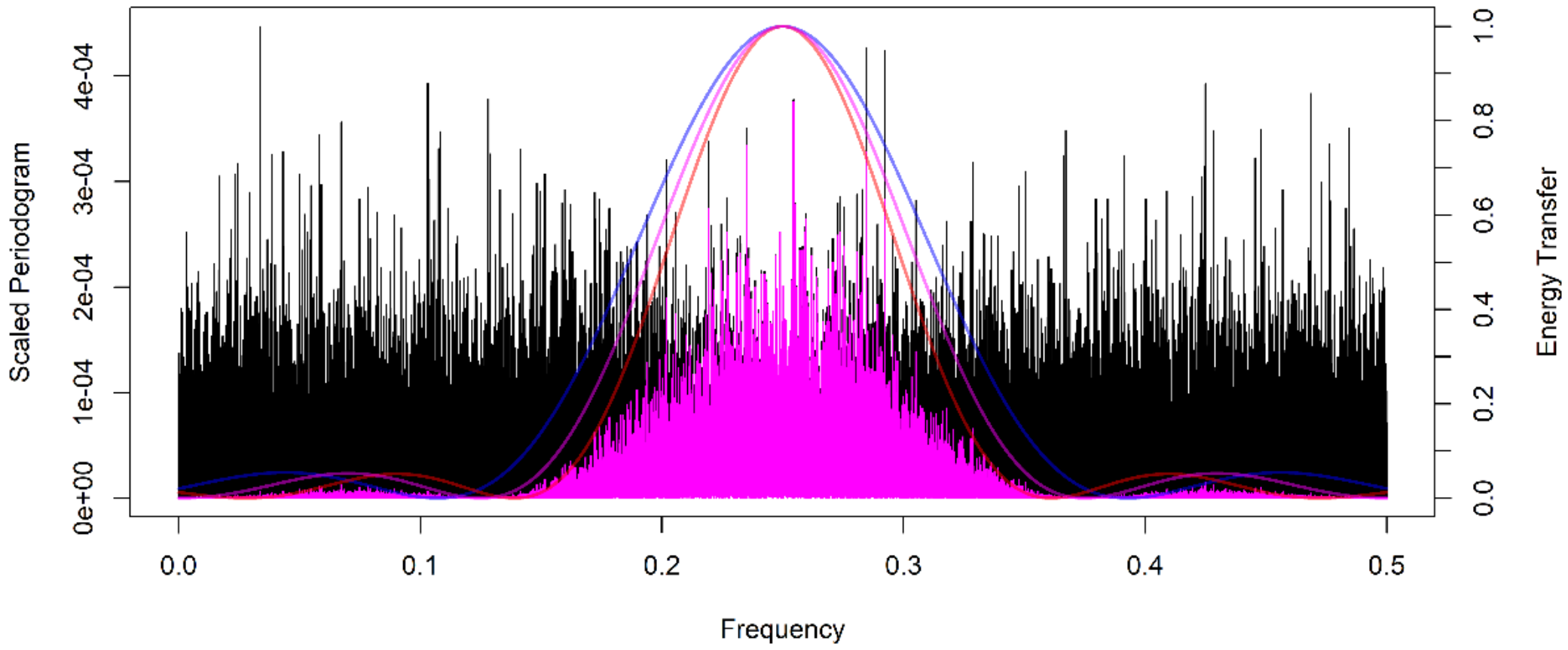

**Fig 4. Example periodograms of a white noise time series and EKZFT filtered time series with even valued filter window length, and corresponding energy transfer functions.** The periodogram of a white noise time series in black and the periodograms of that same time series after applying Extended Kolmogorov-Zurbenko Fourier Transform filters $EKZFT_{m_r=8,k=1,v=0.25}$ in magenta. The energy transfer functions of an $EKZFT_{m_r=8,k=1,v=0.25}$ filter is shown as the magenta curve, and the closest possible $KZFT_{m=7,k=1,v=0.25}$ filter energy transfer function displayed in blue and $KZFT_{m=9,k=1,v=0.25}$ filter energy transfer function displayed in red with corresponding scale to the right.

# Discussion

The KZFT band-pass filter is a useful tool in time series analysis to separate component frequencies, filter interference, reduce noise, and strongly suppress signals outside of a narrow band surrounding a central frequency directly relating to the choice of filter arguments, particularly the filter window size. However, the choice of filter window size limits the KZFT filters because they are only defined for positive odd integer window sizes. This creates a set of filters that has distinct and sometimes large gaps between the choice of spectral width of the band-pass filters. This problem is acutely apparent at the smaller positive odd integer values, and it is at small positive odd integers where the largest differences are observed in available band-pass filter choices as seen in Fig 2. The introduction of the EKZFT filter in this work helps to fill those gaps, making a much more flexible band-pass filter and overcomes some of the shortcomings of the KZFT filters.

A primary advantage of the EKZFT, as previously shown, is that it contains all KZFT filters as special cases, so each of those types of filters and their attributes are inherited in the design of the EKZFT. The next distinct advantage of the EKZFT is that it permits the use of positive even integer window sizes for this class of band-pass filters. This is done while maintaining the feature of these filters to strongly attenuate signals that lie outside a chosen frequency shift away from the central frequency, also while completely suppressing or attenuating signals at that chosen frequency shift or separation, and all multiples of that frequency shift, from the central frequency. With the addition of positive even integer window sizes, the EKZFT already approximately doubles the number of filter designs that are possible compared to KZFT filters while maintaining the same attributes. Finally, when the filter window length is any real number greater than one excluding even and odd integers, the EKZFT filter still supresses frequencies outside of the band created by a frequency shift of $1/m$, but that suppression is incomplete and only approximate to that achieved by KZFT filters. In this case the EKZFT filter still strongly suppresses signals with a frequency shift of $1/m$ or greater, but it no longer completely eliminates signals with a frequency shift of $1/m$ or its harmonics away from the central frequency $v$ selected in the filter argument. For many purposes however, complete suppression is not necessary, and very strong suppression is sufficient. Also, even though EKZFT filters do not always completely suppress signals with a frequency shift of $1/m$ and its harmonics, the level of suppression can be increased though iteration. Consequently, this limitation in the design of the EKZFT filter is minor. Despite the limitation in this case, the EKZFT can still be an effective band-pass filter, and its significant strength is a continuous choice of filter window size. This translates into a band-pass filter with a continuous choice of how wide to define the band of frequencies passed by the filter.

Another advantage of the EKZFT filter, is the relative ease of calculation in its iterative form, a feature it shares with KZFT, EKZ, and KZ filters. In its iterative form the EKZFT is simply a weighted average of observations of the original time series. While the EKZFT can programmed into statistical software, it is also possible to calculate in many simple spreadsheet programs. The KZFT filter has seen extensive use in time series as a band-pass filter for signal reconstruction in the presence of noise and other interference, to separate component signals and to isolate seasonality and other periodic components. While the KZFT may be sufficient to perform some of these functions, the KZFT is a relatively imprecise band-pass filter as compared to the EKZFT and there may be applications where the precision and fine adjustment possible with the EKZFT may be necessary. The VBPBB and VMBPBB are two important examples where the difference between the KZFT and EKZFT are apparent and the greater precision of the EKZFT becomes necessary. For instance, in VMBPBB, it is necessary to separate two component signals using band-pass filters to split those frequencies, passing one and attenuating the other and vice-versa. There are scenarios where this may not be possible with KZFT filters, but the task can be accomplished through the flexibility of the EKZFT. Also, in the real-world examples of applications of these block bootstrap methodologies cited earlier, the KZFT band-pass filters were used to separate the periodic correlated components of the time series [12,13]. While the results obtained were shown to be improvements upon other existing bootstrap methods using KZFT, as explained in describing their design, the VBPBB and VMBPBB performance can be further improved with an optimal choice of band-pass filter size. The EKZFT filter presented in this work now provides just such a band-pass filter with the requisite precise adjustment to obtain optimal results with VBPBB and VMBPBB.

# Conclusion

In time series analysis, the KZFT filter provided a straightforward, efficient, and easily calculable way to define and implement a band-pass filter. The arguments of the KZFT filter defines which frequencies are allowed to pass through the filter, and which are attenuated and how much attenuation is used. However, the KZFT filter does not permit sufficient flexibility in choosing the length of the filter window over which the Fourier transform is applied, preventing precise design of these filters. Since they suppress frequencies that are separated by $1/m$ and farther from the center of the band-pass filter, where $m$ is the chosen filter window but is restricted to positive odd integers, at times they provide what can best be described as an imprecise tool for this task. The problem is most pronounced for small values of $m$, since the corresponding gaps created by the frequency separations of $1/m$ are a substantial portion of the frequency domain. The EKZFT filter proposed here extends the KZFT to permit a continuous selection of filter window lengths across the same range. As a result, the frequency separation specifically chosen for attenuation by the filter is no longer limited to the values of $1/m$, but to any real value greater than one. The EKZFT creates a much more precise tool with many potential future applications.